# Case Studies on Plasma Wakefield Accelerator Design

*J. Osterhoff[1], Z. Najmudin[2] and J. Faure[3]*
[1] Deutsches Elektronen-Synchrotron DESY, Hamburg, Germany
[2] John Adams Institute for Accelerator Science, Imperial College London, UK
[3] LOA, École Polytechnique-ENSTA-CNRS, Palaiseau, France

**Abstract**
The field of plasma-based particle accelerators has seen tremendous progress over the past decade and experienced significant growth in the number of activities. During this process, the involved scientific community has expanded from traditional university-based research and is now encompassing many large research laboratories worldwide, such as BNL, CERN, DESY, KEK, LBNL and SLAC. As a consequence, there is a strong demand for a consolidated effort in education at the intersection of accelerator, laser and plasma physics. The CERN Accelerator School on Plasma Wake Acceleration has been organized as a result of this development. In this paper, we describe the interactive component of this one-week school, which consisted of three case studies to be solved in 11 working groups by the participants of the CERN Accelerator School.

## 1    Introduction

The CERN Accelerator School on Plasma Wake Acceleration was held in Geneva at CERN from November 23 to 29, with the aim of teaching students and postdoctoral researchers the basic concepts of plasma wakefield acceleration at the intersection of accelerator, laser and plasma physics. An important objective of this school was to include an educational component that deviated in form from the lectures given by experts and fostered interaction between participants.

Therefore, three case studies were developed, pertaining to future applications, which required the design of wakefield accelerators for specific purposes. The students were confronted with problems that left a considerable amount of freedom in approaching their solutions. Intentionally, this demanded self-organization of the students within their working group (there were 11 such groups), a break-up of these cases into subproblems and a distribution of tasks within each group. The students had 4 days, in parallel with the lectures, to work out possible case study solutions and prepare one short presentation per group, which were delivered at the end of the school to the students and lecturers.

This paper summarizes the case study activities. The cases are presented in Section 2. Section 3 discusses the educational objectives of this undertaking, Section 4 describes its organization and Section 5 details the outcome and highlights some specific results.

# 2 Case study problems

## 2.1 Case study 1: A plasma-based booster module for the International Linear Collider

### 2.1.1 Introduction

Today the search for new particles and forces at energies of hundreds or thousands of gigaelectronvolts plays a central role in the field of elementary particle physics. Particle physicists have established a Standard Model for the strong, weak and electromagnetic interactions that passes tests at both low and high energies. The model is extremely successful, and yet it is incomplete in many important respects. New particles and interactions are needed to fill the gaps.

Some of the difficulties of the Standard Model are deep and abstract; their explanations may be found only in the distant future. The Standard Model does not explain how gravity is connected to the other forces of nature. It does not explain why the basic particles of matter are the quarks and leptons, or how many of these there should be.

However, the Standard Model also fails to explain three phenomena that, by rights, should be accounted for at the energies now being probed with particle accelerators, e.g., at the LHC at CERN. Astronomers believe that the dominant form of matter in the universe is a neutral weakly interacting species, called dark matter that cannot be composed of any particle present in the Standard Model. The Standard Model cannot explain why the universe contains atomic matter made of electrons, protons and neutrons but no comparable amount of antimatter.

The problem of the Higgs field is likely to be connected to these questions about the matter content of the universe. Explanatory models of the Higgs field often contain particles with the correct properties to make up the dark matter. There are also strong, independent, arguments that the mass of the dark-matter particle is comparable to the masses—of the order of 100 GeV—of the heaviest particles that receive mass from the Higgs field. The predominance of baryons over antibaryons in the universe could arise from interactions among Higgs fields that violate space-time charge-parity symmetry. More generally, any model of fundamental physics at energies above 100 GeV must contain the Higgs field or some generalization and must account for the place of this field within its structure.

A way to prove the existence of the Higgs field and to study its interactions is to find and study the quantum of this field, the Higgs boson. The International Linear Collider (ILC) was designed to study this particle and other new particles that might be associated with it. It provides an ideal setting for detailed exploration of the origin and nature of the Higgs field. In July 2012, the ATLAS and CMS experiments at the CERN Large Hadron Collider announced the discovery of a new particle with a mass of 125 GeV and many properties of the Higgs boson, as postulated in the Standard Model. The ILC will enable the properties of this Higgs boson to be studied in much greater detail than previously possible. (this section was reproduced, with permission, from the *International Linear Collider Technical Design Report* [1].)

### 2.1.2 Task

Assume that the ILC (see Fig. 1) is a 250 GeV centre-of-mass high-luminosity linear electron–positron collider built specifically for the study of the Higgs field, based on 1.3 GHz superconducting radio frequency (RF) accelerating technology. The initial programme of the ILC, to discover a 125 GeV Higgs boson, $H^0$, at this energy, will give the peak cross-section for the reaction $e^+ + e^- \rightarrow Z + H^0$. In this reaction, the identification of a Z boson at the energy appropriate to recoil against the Higgs boson tags the presence of the Higgs boson. At higher energies, other reactions could be studied (see Table 1), which are not accessible in the current design. Wakefield accelerators could offer a cost-efficient path to boost the energy of the ILC and therefore allow for detailed studies of these additional processes.

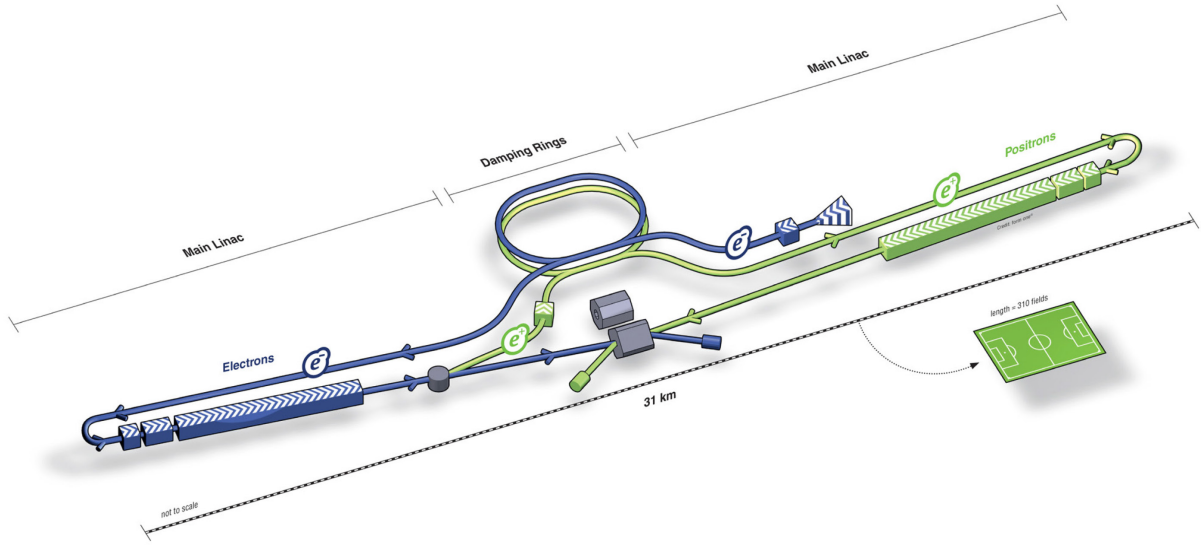

**Fig. 1:** Basic design of the International Linear Collider (taken from Ref. [1])

**Table 1:** Reactions of the Higgs field and other events at the energy scale of interest (cf. Ref. [1])

| Energy | Reaction | Physics Goal |
| --- | --- | --- |
| 91 GeV | $e^+e^- \to Z$ | ultra-precision electroweak |
| 160 GeV | $e^+e^- \to WW$ | ultra-precision $W$ mass |
| 250 GeV | $e^+e^- \to Zh$ | precision Higgs couplings |
| 350–400 GeV | $e^+e^- \to t\bar{t}$ | top quark mass and couplings |
| | $e^+e^- \to WW$ | precision $W$ couplings |
| | $e^+e^- \to \nu\bar{\nu}h$ | precision Higgs couplings |
| 500 GeV | $e^+e^- \to f\bar{f}$ | precision search for $Z'$ |
| | $e^+e^- \to t\bar{t}h$ | Higgs coupling to top |
| | $e^+e^- \to Zhh$ | Higgs self-coupling |
| | $e^+e^- \to \tilde{\chi}\tilde{\chi}$ | search for supersymmetry |
| | $e^+e^- \to AH, H^+H^-$ | search for extended Higgs states |
| 700–1000 GeV | $e^+e^- \to \nu\bar{\nu}hh$ | Higgs self-coupling |
| | $e^+e^- \to \nu\bar{\nu}VV$ | composite Higgs sector |
| | $e^+e^- \to \nu\bar{\nu}t\bar{t}$ | composite Higgs and top |
| | $e^+e^- \to \tilde{t}\tilde{t}^*$ | search for supersymmetry |

Your task is to design a plasma-based energy-booster section for the ILC. The required beam power currently excludes lasers as wakefield drivers. To achieve a realistic design, consider and motivate a beam-driven approach for both the electron and the positron arms of the ILC. Please be careful to conserve the beam properties required in the ILC accelerator section to sustain a suitable event rate (see Table 2).

**Table 2:** Beam parameters of the ILC accelerator (cf. [1])

| | | | 1st Stage |
|---|---|---|---|
| Centre-of-mass energy | $E_{CM}$ | GeV | 250 |
| Collision rate | $f_{rep}$ | Hz | 5 |
| Electron linac rate | $f_{linac}$ | Hz | 10 |
| Number of bunches | $n_b$ | | 1312 |
| Bunch population | $N$ | $\times 10^{10}$ | 2.0 |
| Bunch separation | $\Delta t_b$ | ns | 554 |
| Pulse current | $I_{beam}$ | mA | 5.8 |
| Main linac average gradient | $G_a$ | MV m$^{-1}$ | 31.5 |
| Average total beam power | $P_{beam}$ | MW | 5.9 |
| Estimated AC power | $P_{AC}$ | MW | 129 |
| RMS bunch length | $\sigma_z$ | mm | 0.3 |
| Electron RMS energy spread | $\Delta p/p$ | % | 0.190 |
| Positron RMS energy spread | $\Delta p/p$ | % | 0.152 |
| Electron polarisation | $P_-$ | % | 80 |
| Positron polarisation | $P_+$ | % | 30 |
| Horizontal emittance | $\gamma\epsilon_x$ | μm | 10 |
| Vertical emittance | $\gamma\epsilon_y$ | nm | 35 |
| IP horizontal beta function | $\beta_x^*$ | mm | 13.0 |
| IP vertical beta function | $\beta_y^*$ | mm | 0.41 |
| IP RMS horizontal beam size | $\sigma_x^*$ | nm | 729 |
| IP RMS veritcal beam size | $\sigma_y^*$ | nm | 7.7 |
| Luminosity | $L$ | $\times 10^{34}$ cm$^{-2}$s$^{-1}$ | 0.75 |
| Fraction of luminosity in top 1% | $L_{0.01}/L$ | | 87.1% |
| Average energy loss | $\delta_{BS}$ | | 0.97% |
| Number of pairs per bunch crossing | $N_{pairs}$ | $\times 10^3$ | 62.4 |
| Total pair energy per bunch crossing | $E_{pairs}$ | TeV | 46.5 |

## 2.2 Case study 2: An X-ray free-electron laser based on laser wakefield acceleration

### 2.2.1 Introduction

At gigaelectronvolt energies, the radiation losses of electrons travelling in strong (usually magnetic) fields becomes appreciable. The radiation can easily extend to the X-ray range, as well as being naturally collimated and ultrashort (of the order of femtoseconds). By employing an undulator, a periodically poled magnetic field configuration, this synchrotron radiation can be reinforced in the direction of travel of the electron beam. If the radiated energy is sufficiently bright, it can cause bunching of the electrons, causing them to emit radiation coherently and thus even more strongly. This feedback causes exponential growth of the radiated signal in a device called a free-electron laser (FEL) [2].

To date, two FELs operating at X-ray wavelengths have been demonstrated, the Linac Coherent Light Source at SLAC [3] and SACLA at RIKEN [4], and they are soon to be joined by a handful of others, including the European XFEL at DESY. The enormous increase in the brightness of X-ray sources at these facilities has created dramatic new scientific possibilities, such as the possibility of determining the structure of complex molecules in a single shot, and the ability to generate unique far-from-equilibrium states of ionized matter. However, these facilities are few, and will remain so, primarily because of the exacting requirements on the electron beam to seed an X-ray FEL. The electron beam must have, simultaneously, high energy, high bunch charge, low energy spread and low emittance. This can be seen by the electron beam properties of the existing FEL labs given in Table 3.

**Table 3:** Main parameters of operational and planned X-ray FEL facilities

| Name | Where | $E_{elec}$ (GeV) | $\Delta E/E_{elec}$ | Charge (pC) | $\varepsilon_n$ (μm) | $E_{phot}$ (keV) |
|---|---|---|---|---|---|---|
| LCLS | SLAC (USA) | 14 | << 1% | 250 | 1 | 8 |
| SACLA | J-PARC (Japan) | 8.5 | << 1% | 250 | 0.8 | 10 |
| XFEL | DESY (Germany) | 17.5 | << 1% | 250 | 1.4 | 10 |

Obviously wakefield accelerators could produce high charge beams in a short distance. They are an obvious candidate for generating an electron beam with sufficient energy to drive an X-ray FEL at reduced cost. However, the other parameters required may be harder to produce, and so will require some consideration.

### 2.2.2 Task

In this case study, we will try to design an accelerator to drive an X-ray FEL with a laser wakefield scheme. The first stage will be to determine the plasma characteristics required to achieve the energies needed for the wakefield accelerator.

Once this has been achieved, it is necessary to consider whether this accelerator will be able to provide the necessary charge, with sufficient beam quality. The requirements of the laser and plasma to be able to create this source should be determined. A survey of the literature would allow us to determine how close we are with present experiments and laser systems. Could other wakefield techniques be used to achieve the required characteristics? Are there advances in FEL research that could help us?

## 2.3 Case study 3: A high repetition-rate laser-plasma accelerator for electron diffraction

### 2.3.1 Introduction

One of the current challenges of science is to unravel the dynamical structure of nature on the atomic time-scale; for example, understanding atomic motion during a chemical reaction, or following electron dynamics in a protein after photoexcitation. In the past decade, pulsed sources of X-rays have become available and provide a probe for atomic motion in complex matter at the femtosecond level. For example, the Linac Coherent Light Source at Stanford has reached a high level of performance with the production of unprecedented bright pulses of ultrashort coherent X-rays. In spite of their remarkable capabilities, these sources, based on the use of large electron accelerators, are restricted to a few large facilities with limited beam time access.

In parallel, a less expensive tabletop alternative has been developed, which uses ultrashort electron bunches to probe atomic motion and structural dynamics, either through diffraction [5] or microscopy. In addition to their small size, these electron sources provide another significant advantage in that they offer an elastic scattering cross-section, which is more than five orders of magnitude higher than for X-rays, providing much higher diffraction efficiency. Ultrashort electron bunches have been used successfully for the study of structural dynamics in condensed matter, chemistry and biology. However, the time resolution currently does not exceed 200 fs, preventing the observation of faster phenomena.

Typical electron guns for electron diffraction are d.c. (static acceleration field) photoguns (based on the use of a UV-driven photocathode as the source of electrons) and operate at around 100 keV. Recently, RF guns are being developed, to decrease the pulse duration to below 100 fs. These RF guns operate at a few megaelectronvolts but they are subject to a timing jitter between the laser and the electron pulses. Consequently, reaching <100 fs or even <10 fs is very difficult, even though interesting dynamics occur on these time-scales. To summarize, the current limitations in ultrafast electron diffraction are:

- space charge of the beam, limiting the bunch duration to >100 fs and the charge to femtocoulomb levels (for static acceleration fields);
- synchronization (jitter) between the excitation laser and the electron bunch (for RF fields), limiting the resolution to >100 fs.

Laser-plasma accelerators offer an opportunity to provide even shorter electron pulses for two main reasons: (i) the huge accelerating fields (10–100 GV/m) prevent space charge and (ii) in principle, the electron bunches are jitter-free because they originate from a laser-driven accelerating structure. In this case study, we will examine the possibility of designing a laser-plasma accelerator with the right parameters (see Table 4) and offering <10 fs electron bunches.

**Table 4:** Main parameters of typical electron guns used for electron diffraction. The bottom row lists the parameters that the design should obtain.

| Accelerator | $E_{elec}$ (MeV) | $\Delta E/E_{elec}$ | Charge (fC) | $\varepsilon_n$ (μm) | Bunch duration | Rep. rate |
|---|---|---|---|---|---|---|
| DC photogun [5] | 0.1 | <1% | 1 | $5\times10^{-2}$ | >200 fs | 1 kHz |
| RF photogun [6] | 3.5 | <1% | 100 | $7\times10^{-2}$ | >100 fs | 10 Hz |
| Laser-plasma | 0.1–10 | <1% | 1–1000 | <0.1 | <10 fs | >0.1 kHz |

### 2.3.2 Task

In this case study, we will try to design an accelerator that could be used for electron diffraction applications. The target parameters of the design are quite exotic for a laser-plasma accelerator: note, for example, that for the energy should be quite low (<10 MeV) and that it is preferable that the repetition rate be large (>100 Hz; a kilohertz repetition rate is preferable for good statistics in electron diffraction experiments). The beam quality should be very good, for making clear and usable diffraction patterns; in particular, the transverse emittance needs to be low.

A first step in designing this accelerator would be to determine the necessary plasma parameters, followed by the necessary laser parameters. A survey of the literature would allow us to determine how close we are with present experiments and laser systems (note that first experiments at high repetition rate have begun to produce some results [7]). We should then think about what injection techniques could be used to produce the electron beam. We should also discuss what kind of beam quality we could obtain and how close it is from the target parameters, and suggest ways to improve things. Considerations on bunch duration and elongation upon propagation would also be helpful.

## 3 Educational objectives

The main educational purpose of these case studies was to familiarize the students with advantages, problems and principles of plasma wakefield acceleration. Deliberately, the chosen scenarios were focused on actual visions for applications of plasma-based accelerators, which are currently pursued and studied by a number of research groups worldwide, but remain unsolved in the sense that they have yet to be experimentally realized. Thus, the students were working on cutting-edge problems.

It was intended that these problems should be approached using the knowledge that the students obtained during the lectures at this CERN Accelerator School. Since a number of solution strategies were applicable to each case, each problem was assigned to three or four different groups of students, who produced a variety of outcomes and approaches.

A significant hurdle in the solution process was the initial strategic discussion and distribution of tasks within each working group. This was intentional. The participants had to interact and self-organize as peers to come up with a successful strategy, strengthening their networking, social interaction and group performance skills. The collective work on a specific problem turned out to be highly motivating for many students. Many of them spent hours following different solution paths, experimenting with different parameters and optimizing their base designs, using every free minute they had between lectures and after dinner, and often working until late at night.

## 4     Case study organization

On the first day, the three case studies were distributed to three (case study 1) or four different groups (case studies 2 and 3), resulting in 11 working groups altogether. Each group consisted of up to 10 students with backgrounds in accelerator, laser or plasma physics, who worked jointly on the assigned tasks.

The case study work itself was supposed to be done after the end of the lectures in the evenings and during two tutorial sessions (of at least an hour each) on days two and three of the school. For this purpose, three separate seminar rooms were booked with two tutors each. The tutors were selected from the list of lecturers. In these tutorials, students were given the chance to discuss the problems and their solutions with experts.

On the final day of lectures, the students were given an extra hour to prepare their concluding presentations, to be delivered at the end of that day. These presentations were 5 minute breakdowns of the achieved results and suggested designs and were presented to students and lecturers in the audience and a jury, who selected winners for the categories 'best overall solution', 'most innovative solution' and 'most entertaining presentation'.

## 5     Outcome, highlights and conclusions

The extremely active, productive and creative atmosphere during the case study work that was prevalent during the week resulted in a number of highly interesting solutions and innovative schemes suggested by the different working groups, making the case study programme a great success. In the following, a number of these proposals and highlights are discussed for each case study.

### 5.1     Case study 1: A plasma-based booster module for the International Linear Collider

The three groups working on this case proposed various design solutions at different plasma densities from $10^{15}$ cm$^{-3}$ to $10^{16}$ cm$^{-3}$. All the groups suggested reshaping the ILC bunch train structure, such that odd and even numbered bunches from the train could be utilized as drivers and witnesses, respectively, in a beam-driven plasma wakefield accelerator set-up. The main issue, which was correctly identified, is the conservation of beam properties in the plasma booster stage, in particular in the positron arm of the ILC. The three proposals differed significantly, proposing the use of hollow-core plasma channels [8–10], plasma acceleration in the linear regime and even the use of plasmas consisting of antihydrogen.

### 5.2     Case study 2: An X-ray free-electron laser based on laser wakefield acceleration

This case study was considered by four groups, who came up with different design ideas, in particular with respect to the utilized undulator technology. The proposed methods covered cryogenic undulators [11], transverse gradient undulators [12], synchronized seeding from solid-target high harmonic sources [13] and travelling wave Thomson scattering [14]. Beam transport was also taken into account, with an emphasis on beam emittance preservation and the inclusion of a plasma-based beam dechirper. Proposed electron injection techniques within the plasma target ranged from ionization injection [15, 16] to counter-propagating pulse injection [17].

### 5.3 Case study 3: A high repetition-rate laser-plasma accelerator for electron diffraction

The four working groups considered the creation of suitable low-emittance and low-energy-spread beams from a plasma to be challenging and therefore presented solutions using various controlled injection methods and a sophisticated electron beam transport system. Electron beams were generated by colliding few-cycle laser pulses or by shock-front injection [18]. In some cases, the transport system also included a post-plasma energy filter for the generation of well-defined narrow-bandwidth spectra.